# Carbogen inhalation during Non-Convulsive Status Epilepticus: A quantitative analysis of EEG recordings

*S Ramaraju[1a], S Reichert[2], Y. Wang[1], R Forsyth[3*], P N Taylor[1*]*

1. Interdisciplinary Computing and Complex Bio-Systems Group, School of Computing Science, Newcastle University, UK
2. Department of Electrical Engineering, Medical Engineering and Computer Science, Offenburg University of Applied Sciences, Germany
3. Institute of Neuroscience, Faculty of Medical Science, Newcastle University, UK

\* Joint Senior Authors

a Corresponding Author : sriharsha.ramaraju@newcastle.ac.uk

## Highlights

- We investigated the effect of inhaled carbogen on EEG recordings from five paediatric NCSE patients.
- The quantified results of EEG indicate non-homogeneous effect of carbogen.
- The non-homogeneous effects are may be due to the varied neural disabilities across patients.

## Abstract

**Objective:** To quantify the effect of inhaled 5% carbon-dioxide/95% oxygen on EEG recordings from patients in non-convulsive status epilepticus (NCSE).

**Methods:** Five children of mixed aetiology in NCSE were given high flow of inhaled carbogen (5% carbon dioxide/95% oxygen) using a face mask for maximum 120s. EEG was recorded concurrently in all patients. The effects of inhaled carbogen on patient EEG recordings were investigated using band-power, functional connectivity and graph theory measures. Carbogen effect was quantified by measuring effect size (Cohen's d) between "before", "during" and "after" carbogen delivery states.

**Results:** Carbogen's apparent effect on EEG band-power and network metrics across all patients for "before-during" and "before-after" inhalation comparisons was inconsistent across the five patients.

**Conclusion:** The changes in different measures suggest a potentially non-homogeneous effect of carbogen on the patients' EEG. Different aetiology and duration of the inhalation may underlie these non-homogeneous effects. Tuning the carbogen parameters (such as ratio between $CO_2$ and $O_2$, duration of inhalation) on a personalised basis may improve seizure suppression in future.

**Keywords:** Carbogen, EEG, Epilepsy, Non Convulsive Status Epilepticus, Functional Connectivity

**Introduction**

Status epilepticus (SE) is a situation of continuing seizure activity or repetitive seizures without recovery lasting (by convention) over 30 minutes (DeLorenzo et al., 1998). Morbidity and mortality are affected by factors including age, aetiology and time to first treatment (Lowenstein, Alldredge 1993, Towne et al., 1994). Non-Convulsive Status Epilepticus (NCSE) is a subtype of SE again of varied aetiology (Maganti et al., 2008), and is characterised by more subtle SE without prominent motor signs but generally reduced awareness of surroundings and responsiveness (Walker 2007). NCSE is often observed in the context of pre-existing neurological conditions such as injury and neurogenetic syndromes associated with severe epilepsy or learning difficulties (Galimi 2012). NCSE should be suspected in children with epilepsy who undergo an otherwise unexplained deterioration in behaviour, speech, memory, or school performance (Galimi 2012).

Many treatments for convulsive SE have anaesthetic or sedative properties. This is disadvantageous in NCSE as sedation can activate NCSE and children are at increased prior risk of respiratory problems (Forsyth et al., 2016). The ideal treatment in this case would be immediate acting, non-drowsy, maintaining respiratory function and acting for a sustained period. Tolner et al., (2011) showed that induction of temporary mild respiratory acidosis (reduction of pH levels) by inhaling carbogen (5% carbon dioxide and 95% oxygen) has anticonvulsant action, and this would be of potential value in NCSE. Carbogen has shown to terminate acute onset seizures in rats, non-human primates and adult humans (Lennox et al., 1936, Pollock et al., 1949, Pollock 1949, Dahlberg-Parrow 1951, Woodbury et al., 1955, Meyer JS 1961, Ziemann et al., 2008).

During NSCE, electroencephalographic (EEG) activity is abnormal. Patient EEG dynamics can include pathological polyspike & slow waves, generalised slowing, and burst-suppression (Sutter, Kaplan 2012). The effect of carbogen on NCSE EEG is not well understood, however. To quantify changes in EEG dynamics, several methods are available. Band-power analysis shows EEG signal power in particular frequency bands, and functional connectivity – the inference of brain networks by measuring signal similarity - has been used widely in epilepsy to show cortical network organisation before, during, and after seizures (Burianova et al., 2017, Tracy, Doucet 2015, Abreu et al., 2019, Beniczky et al., 2013, Sutter, Kaplan 2012, Kaplan 2007). Graph theory metrics can further be applied to functional networks to assist interpretation, and elucidate specific aspects of the network. For example, clustering coefficient and path length are two graph theory measures commonly used to quantify local and global properties of a network (Bullmore, Sporns 2009). These two measures have also been shown to vary over the course of a seizure (Schindler et al., 2008).

In this study we investigated the effect of carbogen on band power and functional connectivity across five frequency sub-bands (delta, theta, alpha, beta and gamma).

## 2. Methods

The methods are organized in five main sections: (1) patient information and the corresponding EEG recordings, (2) data pre-processing, (3) band-power analysis, (4) graph-theoretical measures, and (5) statistical analysis. Fig. 1 summarises the whole analysis pipeline.

### 2.1 Patient information and recordings

We retrospectively studied five NCSE patients for whom EEG data was acquired in 2016 across three different locations and a detailed description is provided in (Forsyth et al., 2016). Patients selected for this study had all been diagnosed with epilepsy with either known aetiologies or different underlying neurodevelopmental disabilities. The mean age of the sample is 7± 3.85 years (mean ± standard deviation), of which four patients were male. Children in NCSE were given high flow inhaled carbogen by face mask for maximum 120s (113s±7s) with concurrent EEG measurement. All recordings were performed with a standard 10-20 clinical recording system (Forsyth et al., 2016). The clinical and demographic information of all patients in this study is summarised in Table 1.

| Patient | Gender | Age [yrs] | Aetiology | Number of Electrodes | Sampling rate [Hz] | Pre-inhalation EEG-Duration [s] | Post-inhalation EEG-Duration [s] |
|---|---|---|---|---|---|---|---|
| **1** | M | 4 | Angelman-Syndrom | 20 | 500 | 1200 | 1158 |
| **2** | F | 10 | Lissencephaly | 19 | 256 | 1200 | 979 |
| **3** | M | 3 | Lissencephaly (PAFAH1B1 mutation) | 20 | 256 | 416 | 1128 |
| **4** | M | 13 | Alper mitochondrial depletion syndrom (POLG1 mutation) | 20 | 256 | 989 | 1532 |
| **5** | M | 5 | Angelman | 20 | 256 | 1200 | 1224 |

**Table 1** Clinical and demographic information of all patients

Patient EEG activity was recorded continuously 1001.08 ± 303.55s pre-inhalation, during inhalation, and 1204.3 ± 182.45s post carbogen inhalation. A provision in the protocol allowed for the premature discontinuation of inhalation if the clinicians (including parents) felt it was any distress (Forsyth et al., 2016). The data were collected under to ethical guidelines and protocols were monitored by a responsible clinician.

### 2.2 Data pre-processing:

The data was notched at 50Hz (to exclude power line interference), and band-pass filtered between 1 to 70Hz using forward and backward $2_{nd}$ order Butterworth filter. Data was visually inspected for amplifier saturations or noisy channels; the eye blink-artefacts were rejected using Independent Component Analysis (ICA). For data analysis we extracted three epochs from the EEG data of every patient: "before" (immediately before inhalation of carbogen), "during" (during inhalation of carbogen) and "after" (immediately after inhalation of carbogen). The length of each epoch in "before" and "after" state is 120s. The length of epoch in "during" state varies between 106s-120 across the patients.

## 2.3 Band-power analysis:

A two second sliding window with 50% overlap was used to extract absolute spectral power in five different frequency bands (delta: 1-4Hz, theta: 4-8 Hz, alpha: 8-13 Hz, beta: 13-30 Hz, and gamma: 30-70Hz) across each channel. The mean absolute powers (and the standard deviations) in each sliding window were averaged across windows for each frequency band, channel and patient. This gives five features per channel across three different states (before, during, and after) for each patient. This is visually summarised in Fig 1 (A-C). The spatial average time series (average over all the channels) in broadband (1-70Hz) has also been plotted in Fig 2 across three states to visualise the effect of carbogen over time. The data was smoothed using a moving median filter (n = 31), to minimize the fluctuation of the signal for visualisation purpose, however, all the statistical calculations were carried out on original non-smoothed signal (Fig S1 in supplementary materials).

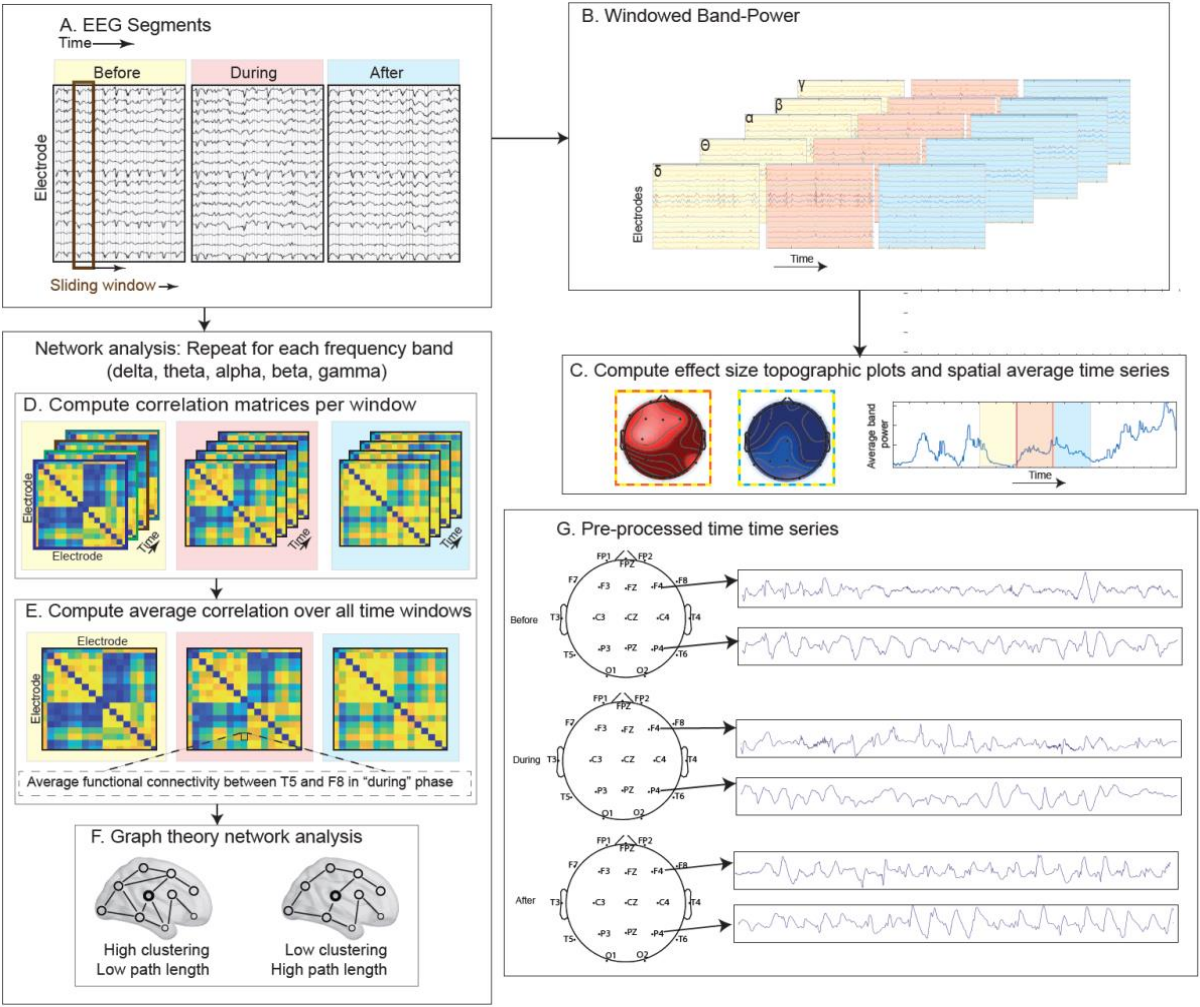

**Fig 1: Analysis Pipeline (A-F)**
(A). Pre-processed EEG signal in "before", "during" and "after" states (B). Band-power computed across all the channels for every sliding window (C). Windowed band-power across all the channels used for topographic and spatial average time series plots. (D-E) windowed correlations matrices for every sliding window and then averaged over time (F). Illustrative networks demonstrating path length and clustering coefficient. (G) Pre-processed EEG time series for two example locations for Patient 3.

### 2.4 Network construction & graph theory analysis

The same EEG segments (before, during, after) used in band-power analysis were also used here to calculate the functional connectivity (amplitude correlation) and graph theory measures. The data has been bandpass filtered in the above-mentioned frequency bands followed by a sliding window analysis, as in band-power analysis. Each window resulted in a functional connectivity matrix (absolute Pearson's correlation matrix and diagonal correlation values were set to zero) through which graph theory measures (clustering coefficient and path length) were calculated using Brain Connectivity Toolbox (Rubinov, Sporns 2010). This is visually depicted in Fig 1 (D-F). The spatial average broadband functional connectivity, path length and clustering coefficient time series have been plotted in Fig 3 across three states to visualise the effect of carbogen across time.

The data was smoothed using a moving median filter (n = 31), to minimize the fluctuation of the signal for visualisation purpose, however, all the statistical calculations were carried out on original non-smoothed signal which can be found in Fig S2-S4 in supplementary materials.

### 2.5 Statistical analysis

The effect size (Cohen's *d*) comparing the states "before" to "during", and "before" to "after" across all the channels and frequency bands was plotted in topographical plots in Fig 2. Additionally, we calculated the actual percentage change (equation (1)) of the band-power from "before" to "during" and "before" to "after" states across all the electrodes for each patient in each frequency band. The percent changes are summarised in Fig S5 in supplementary material.

$$C = \frac{(BPpost - BPpre)}{(BPpost + BPpre)} \quad (1)$$

BPpost is band-power in "during" or "after" states.

BPpre is band-power in "before" state.

The effect sizes were also calculated for the functional connectivity in each entry of the matrix. Only medium and large effect sizes (threshold >=0.5) are displayed in Fig 3. The time varying net broadband functional connectivity (mean of connectivity matrix across every sliding window) is plotted in Fig 3.

Permutation test (10,000 permutations; mean) and Cohen's *d* are used to quantify the effect of carbogen on patient's EEG measures (band-power and functional connectivity).

## 3 Results:

The results section is divided into two sections. First, the band-power analysis and second, the functional network analysis. For brevity, two example patients' results are presented in full, with the remaining in the supplementary material.

### 3.1 Band Power Analysis:

The effect size of the band-power across all channels in each frequency sub-band for each patient in the "before-during" and "before-after" comparison is summarised in Fig 2 (A, B, D, E, G- L). The spatial average broadband time series and effect sizes show no clear or consistent pattern across all patients. In the "before-during" comparison, a positive effect indicates decrement of absolute band-power in "during" state relative to "before" state, whilst an increment indicates the opposite. The above statement also applies to the "before-after" comparison.

Figure 2C shows the net (spatial average) broadband time series for Patient 3. The effect sizes of "before-during" and "before-after" comparisons indicate small ($d$=0.32, $p$=0.03; $d$=-0.17, $p$=0.30) effects. The effect sizes in "before-during" comparisons (Fig 2A) indicate a small effect across all the frequency bands. The same is observed in the "before-after" comparison with an exception in the gamma band. The effect size and $p$-values across every channel for "before-during" and "before-after" comparisons are summarised in Table S5 and Table S6 in supplementary materials. This result indicates a small effect of carbogen on the EEG band-power of Patient 3.

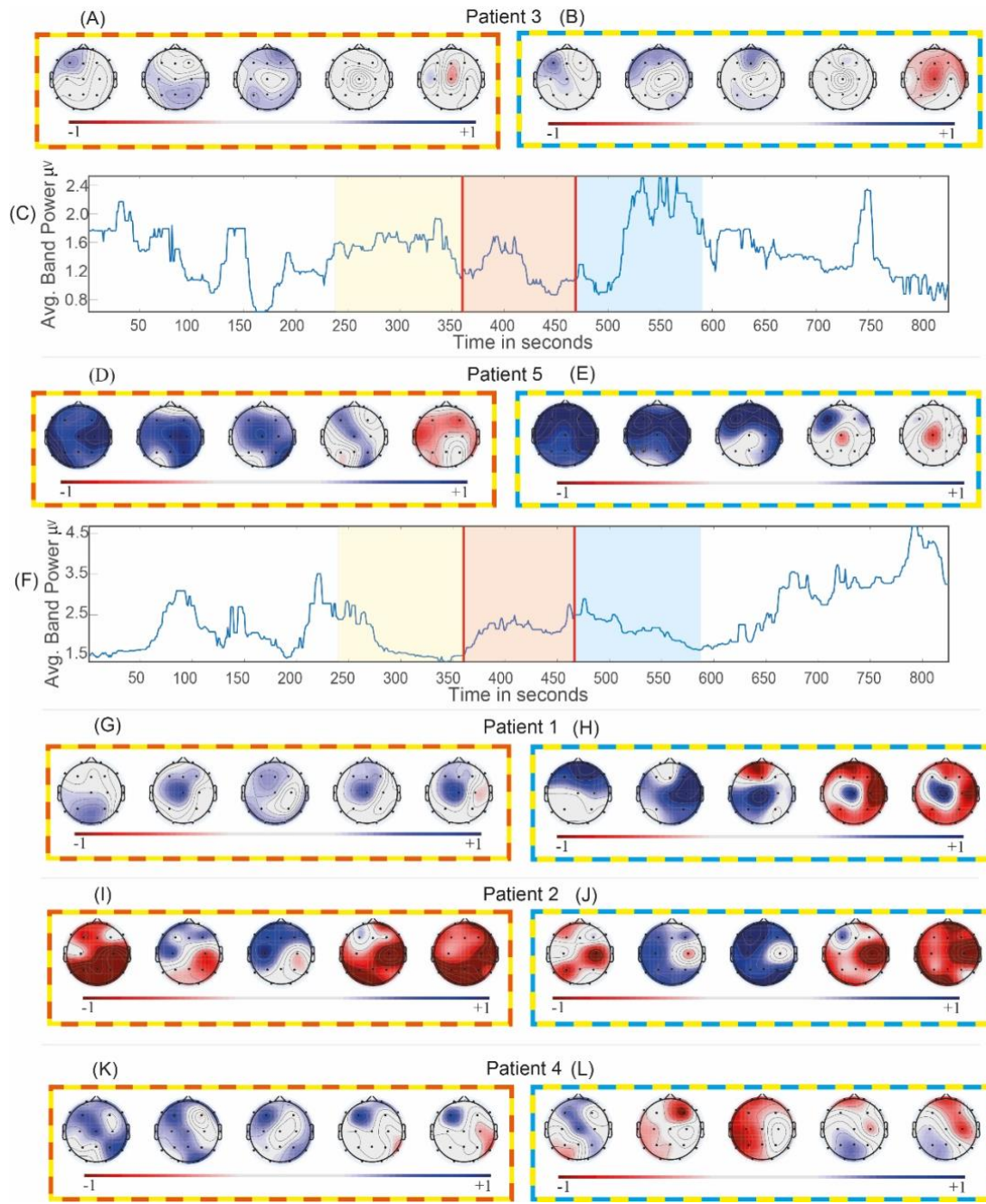

**Fig 2: Band-power analysis**
(A, D, G, I, K) Effect sizes (Cohen's *d*) of the band power change of "before-during" and (B, E, H, J, L) "before-after" across five frequency bands. (C, F) Time series of the average band-power in broadband (1-70Hz) for Patient 3 and Patient 5 respectively. Positive effect size denotes decrement in absolute band power.

In contrast to Patient 3, large effects are observed in "before-during" and "before-after" comparisons for the EEG in Patient 5 (summarised in plots in Fig 2D, E). These effect sizes indicate the suppression of band-power especially in the lower frequency bands during and after carbogen inhalation. The effect size and *p*-values are summarised in Table S9 and Table S10 in supplementary materials. The suppression of band-power is visible in the sub-bands (Fig 2 D and E), however, it cannot be observed in the average broadband time series in Fig 2F ("before-during": *d*=-0.14; *p*=0.47 and "before-after": *d*=0.01; *p*=0.96).

The two representative patients: Patient 3 and Patient 5 indicates contrasting effects of carbogen across all the sub-bands and broadband. This non-homogenous effect of the carbogen on absolute band-powers is also observable across three other patients (Fig 2 (G)-(L)). The net broadband time series (for the remaining patients; Fig S2 F-H) and normalised sub-band time series (for all the patients; A-E) can be found in Fig S6 in supplementary materials. Supplementary Tables S11-S12 and Tables S1-S10 contain the effect sizes and permutation test $p$-values for broadband time series and individual channels across all the sub-bands respectively.

### 3.2 Network Analysis

#### 3.2.1 Functional connectivity

In Patient 3 (Fig 3 A-B), the functional connectivity in the "before-during" comparison does not show any medium or large effects ($d >= 0.5$) of the carbogen in the delta, theta or alpha sub-bands. Beta and gamma bands show reduction of the connectivity between few nodes. In the "before-after" (Fig 3B) comparison, carbogen shows only a small effect in the gamma band. The insignificance ($d=0.03$, $p=0.82$) of carbogen inhalation in "during" state, and significant effect on post inhalation ($d=0.31$, $p=0.03$) can also be seen in Fig 3C across the broadband time series.

Patient 5 on the other hand, shows a clear suppression in the net broadband functional connectivity when comparing "during" to "after" states (Fig 3H). This indicates a significant effect during carbogen inhalation ($d=0.82$, $p<0.001$) and a small but significant effect in the "after" inhalation state ($d=0.42$; $p<0.001$). The topographical plots in patient 5 (Fig 3(F)-(G)) also show contrasting results in comparison to Patient 3. In Fig 3(F), the functional connectivity is suppressed in lower frequencies during carbogen inhalation, however, alpha and beta bands are less affected and also show some increment in connectivity. In the higher frequencies (gamma), the predominant effect of elevation in the connectivity between most nodes can be observed in the "before-during" comparison. The same observations can be made in the "before-after" comparison (Fig 3(G)).

The results of functional connectivity analysis of the other three patients also suggests this non-homogenous influence of carbogen on the EEG. The sub-band functional connectivity (Fig S7 A-E) and broadband time series (Fig S7 F-H) results for the remaining subjects are summarised in supplementary materials. The effect size and permutation test $p$-values for the remaining subjects can be found in Table S11 and Table S12 respectively in supplementary materials.

#### 3.2.2 Graph theory

The path length and average clustering coefficient in broadband for two representative patients is plotted as a time series in Fig 3 (D, E, I, J). In Patient 3, a significant ($d=-0.36$, $p=0.01$) effect is observed in the "after" states (Fig 3D) in path length only. The net clustering coefficient also exhibits the similar property as path length.

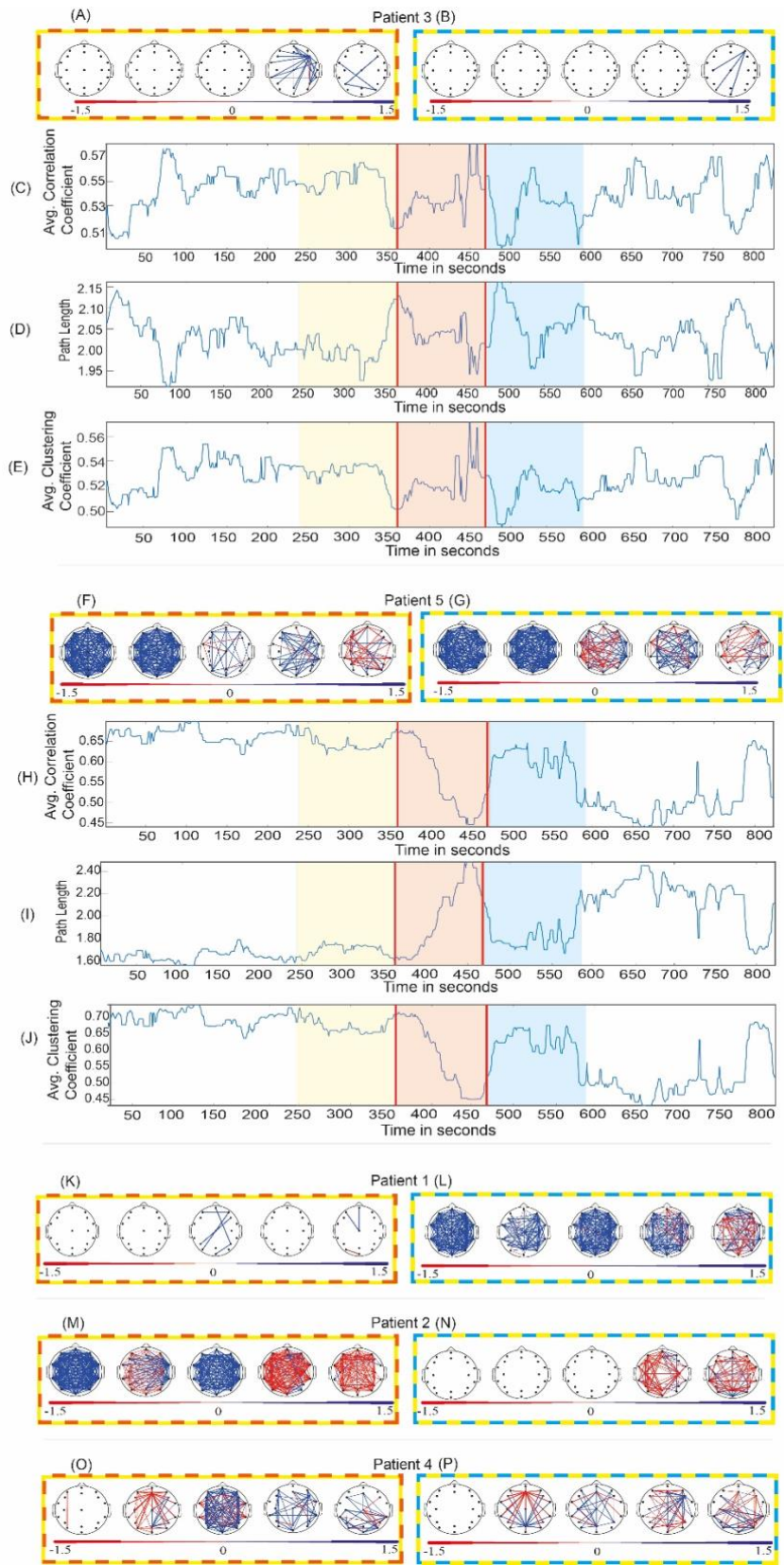

**Fig 3: Functional connectivity and graph-theory analysis**
(A, F, K, M, O) Effect sizes (Cohen's *d*) of the functional connectivity change of "before-during" and (B, G, L, N, P) "before-after" for five Patients. (C, H) Time series of net (average) correlation coefficient in broadband (1-70Hz) for Patient 3 and Patient 5. (D, I) Time series of the net (average) path length in broadband (1-70Hz) for Patient 3 and Patient 5. (E, J) Time series of the net (average) clustering coefficient in broadband (1-70Hz) for Patient 3 and Patient 5. Positive effect size denotes decrement in functional connectivity.

In contrast to Patient 3, in Patient 5 (Fig 3I), the net path length has increased significantly in "during" ($d=-0.88$, $p=0.00$) and "after" states ($d=-0.41$, $p=0.004$; Fig. 3I). However, clustering coefficient has been significantly supressed "during" ($d=0.81$, $p=0.00$) and "after" ($d=0.39$, $p=0.007$) inhalation of carbogen (Fig 3J). The results for the remaining subjects also show inconsistency in the effect of carbogen amongst them. The sub-band (Fig S8 A-E) and broadband path length time series (Fig S8 F-H) for the remaining subjects are summarised in supplementary materials. The sub-band (Fig S9 A-E) and broadband clustering coefficient time series (Fig S9 F-H) results for the remaining subjects are summarised in supplementary materials.

**Discussion**

The present work investigates the relationship between carbogen inhalation and quantitative EEG measures. Specifically, we compared spectral power, functional connectivity and graph theory measures "before" to "during" and "after" the inhalation of carbogen in paediatric NCSE. For these methods we did not find evidence to suggest a common effect of carbogen on NCSE EEG.

Lennox (Lennox et al., 1936) applied 10% $CO_2$ to successfully supress the spike wave EEG activity. This was extended in several studies on rodent, canine and non-human primate models (Pollock et al., 1949, Pollock 1949, Dahlberg-Parrow 1951, Mitchell, Grubbs 1956, Balestrino, Somjen 1988, Ziemann et al., 2008). 15-30% $CO_2$ prevented electrically induced convulsions in psychiatric patients (Pollock et al., 1949). Tolner et al (Tolner et al., 2011) in their pilot study on seven patients with drug resistant epilepsy used standard medical carbogen (5% $CO_2$) and reported the rapid termination of electrographic seizures despite the fact that the application of carbogen was started after the seizure generalization. In the above study, none of the patients had any underlying neural disability except "epilepsy". This is important to note because the cohort in our study has mixed aetiology.

The inhalation of carbogen induces a mild temporary acidosis (i.e. lowering pH) and when reduced to a critical value (blood pH-level<7.35) this is termed as acidaemia (Hamilton et al., 2017). Acidosis attenuates excitatory neurotransmission by reducing NMDA-receptor activity (Brodie, Sills 2011) whilst enhancing inhibitory neurotransmission by facilitating the responsiveness of GABAA receptors (Brodie, Sills 2011). Possible cellular mechanisms (Chesler 2003) involve direct effects of pH on voltage and ligand-gated ion channel conductance (Traynelis et al., 1995, Pasternack et al., 1996, Ziemann et al., 2008) and adenosine signalling (Dulla et al., 2009). Therefore, one may expect a decrease in the EEG band-power, which was also reported previously (Bloch-Salisbury et al., 2000). However, we found an increase in band-power during carbogen delivery in Patient 5 (Fig 2F).

In Patient 3, no substantial effects were observed in "during" and "after" states relative to the "before" state. However, in Patient 5, carbogen delivery was associated with reduce correlation between the nodes (i.e. positive effect size) in lower frequency bands. This may be due to the fact that acidosis introduced by the carbogen might not be sufficient enough to supress the high frequency activity. A small, and insignificant change of path length and clustering coefficient in during state, can be observed in Patient 3. This might be due to the fact that carbogen was inhaled not at the seizure onset but very late into the seizure causing a delayed acidosis effect. This delayed effect might not be enough to terminate the seizure state and show change in EEG recordings. Application of carbogen immediately after the seizure onset might have strong anticonvulsant and therapeutic interventions (Tolner et al., 2011).

Overall the effect of the carbogen on patient EEG was inconsistent across all measurements used. Note that although in our analysis no measure showed a consistent effect, this does not prohibit the possibility of the existence of a measure showing consistent effects across patients. In other words, absence of evidence does not necessarily represent evidence of absence and our finding should therefore be considered with this in mind. Our choices of measures are routine in the field of quantitative EEG analysis and have been shown to demonstrate differences in a wide range of settings (Douw et al., 2010b, Douw et al., 2010a, Bullmore, Sporns 2009, Guo et al., 2010, Lee et al., 2017). Other factors for the ineffectiveness of carbogen might be, first, the lack of "aggressiveness" (concentration/duration) and second, the heterogeneity of the underlying aetiology of the patients.

Even though carbogen has the potential to terminate NCSE seizures in humans, our results suggest different effects in different subjects. This heterogeneity may be related to underlying aetiology, which should be considered in future studies.

**Conflict of Interest**

None of the authors has potential conflicts of interest to be disclosed.

**Acknowledgement**

S Ramaraju was supported by Wellcome Trust grant (210109/Z/18/Z).